\documentclass[
prl,
superscriptaddress,
noshowpacs,
onecolumn,
preprint,
12pt,
titlepage,
floatfix
]
{revtex4-2}
\usepackage{graphicx}
\usepackage[colorlinks,
            linkcolor=blue,
            citecolor=blue,
            anchorcolor=blue,
            urlcolor=blue,
            ]{hyperref}
\usepackage{leftidx}
\usepackage{subfigure}
\usepackage{amsmath}
\usepackage{amssymb}
\usepackage{epstopdf}
\usepackage{amsmath}
\usepackage{lineno}

\newlength{\onecolfig}
\newlength{\twocolfig}
\newcommand{\unit}[1]{\,\mbox{#1}}

\newcommand{\kg}{\unit{kg}}
\newcommand{\kHz}{\unit{kHz}}
\newcommand{\MHz}{\unit{MHz}}
\newcommand{\GHz}{\unit{GHz}}

\newcommand{\urad}{\unit{$\mu$rad}}
\newcommand{\mm}{\unit{mm}}
\newcommand{\um}{\unit{$\mu$m}}

\newcommand{\ms}{\unit{ms}}
\newcommand{\us}{\unit{$\mu$s}}

\newcommand{\dB}{\unit{dB}}
\newcommand{\dBm}{\unit{dBm}}

\newcommand{\etal}{{\em et al.}}

\begin{document}
\title{Optimizing Resonator Frequency Stability in Flip-Chip Architectures: A Novel Experimental Design Approach}
\author{Yuan Li}
\thanks{These two authors contributed equally to this work.}
\affiliation{Tencent Quantum Laboratory, Tencent, Shenzhen, Guangdong 518057, China}
\author{Tianhui Wang}
\thanks{These two authors contributed equally to this work.}
\affiliation{Tencent Quantum Laboratory, Tencent, Shenzhen, Guangdong 518057, China}
\author{Jingjing Hu}
\affiliation{Tencent Quantum Laboratory, Tencent, Shenzhen, Guangdong 518057, China}
\author{Dengfeng Li}
\affiliation{Tencent Quantum Laboratory, Tencent, Shenzhen, Guangdong 518057, China}
\author{Shuoming An}
\email{shuomingan@tencent.com}
\affiliation{Tencent Quantum Laboratory, Tencent, Shenzhen, Guangdong 518057, China}
\date{December 11, 2023}
\begin{abstract}
In multi-qubit superconducting systems utilizing flip-chip technology,  achieving high accuracy in resonator frequencies is of paramount importance, particularly when multiple resonators share a common Purcell filter with restricted bandwidth. 
Nevertheless, variations in inter-chip spacing can considerably influence these frequencies. 
To tackle this issue, we present and experimentally validate the effectiveness of a resonator design.
In our design, we etch portions of the metal on the bottom chip that faces the resonator structure on the top chip. 
This enhanced design substantially improves frequency stability by a factor of over 3.5 compared to the non-optimized design, as evaluated by the root mean square error of a linear fitting of the observed frequency distribution, which is intended to be linear. 
This advancement is crucial for successful scale-up and achievement of high-fidelity quantum operations.
\end{abstract}
\maketitle
\section{Introduction}
Quantum computing possesses extraordinary transformative potential across numerous fields of science and technology. 
Effectively harnessing an increasing number of qubits is crucial for fully realizing quantum computing's potential, ranging from near-term applications that can provide valuable insights despite noise~\cite{kim2023evidence}, to the ultimate achievement of fault-tolerant quantum systems that can overcome errors and deliver groundbreaking computational capabilities~\cite{fowler2012surface}. 
Superconducting qubits have emerged as one of the leading platforms~\cite{ladd2010quantum}, necessitating innovative techniques to accommodate the ever-growing qubit count~\cite{bravyi2022future}. 
One of these techniques is the flip-chip bonding, which connects two chips face-to-face via superconducting metal bumps, typically indium, due to its ductility and facile cold-welding properties.
This technique is gaining increasing importance in the development of superconducting qubit chips~\cite{foxen2017qubit,rosenberg20173d,o2017superconducting,conner2021superconducting,li2021vacuum,kosen2022building,somoroff2023flip}, as it effectively addresses wiring challenges and enables the continuous expansion of qubit numbers, thereby paving the way for more advanced and powerful quantum computing systems. 
 
In recent years, researchers have made significant progress in improving the reliability and accuracy of the flip-chip bonding technique~\cite{niedzielski2019silicon,li2021vacuum,norris2023improved}. 
However, several challenges persist, such as controlling chip spacing, ensuring chip planarity,  maintaining alignment accuracy, and ensuring material compatibility~\cite{kosen2022building}.
Particularly, the inter-chip spacing directly affects various aspects of the device's performance, including resonator and qubit frequencies~\cite{pozar2011microwave}, the coupling rate between quantum components~\cite{gold2021entanglement}, and the impedance of signal lines~\cite{simons2004coplanar}. 
Furthermore, the off-target resonator frequency can introduce readout crosstalk~\cite{duan2021mitigating}, thereby reducing readout fidelity, or cause an additional phase error after a mid-circuit readout~\cite{rudinger2022characterizing}.

The inter-chip spacing typically faces three primary challenges: overall offset, relative tilt, and chip flatness. 
To address these challenges, several fabrication-side approaches have been developed. 
One such a approach is the use of silicon spacers~\cite{niedzielski2019silicon}, which can reduce chip tilt to approximately $11 \urad$. 
Another approach is the implementation of large indium bumps~\cite{li2021vacuum}, which can keep deviations less than 0.5\um\ for a 5\um\ spacing. 
Additionally, incorporating a thick photoresist as hard spacers~\cite{satzinger2019simple} can result in deviations of about 0.8\um\ for a 9.6\um\ spacing. 
These approaches aim to ensure accurate and consistent inter-chip spacings during the flip-chip bonding process.
However, some imperfections still can not be fully resolved by these fabrication improvements.
Such as the bow of the wafer~\cite{takasu1981wafer}, and the hard spacer height, which may exhibit a $4 \%$ off-target deviation~\cite{niedzielski2019silicon}.
These imperfections can easily cause the resonator frequency of about $6.7 \GHz$ to shift by tens of $\MHz$, which in turn increases the readout crosstalk and affects the resonator allocation in the bandwidth of Purcell filters~\cite{jeffrey2014fast} or quantum-limited parametric amplifiers~\cite{mutus2014strong} within a multi-qubit architecture. 
As the scale of qubit systems continues to grow, further optimization from the design aspect is still necessary to maintain precise control over the inter-chip spacing. 

In this work, we present and evaluate a design-side approach aimed at mitigating the impact of the inter-chip spacing variability on the frequency of resonators in flip-chip configurations, as concurrently proposed in theory by Li~\etal~\cite{Li2023-ez}. 
Our design includes a robust resonator that can withstand a substantial range of inter-chip spacing fluctuations. 
To validate the effectiveness of our design, we performed experimental tests on an intentionally tilted large-scale flip-chip setup. 
Our results reveal that we can maintain a precise resonator frequency even with significant inter-chip spacing variations, thereby laying the groundwork for large-scale, high-performance quantum computing chips~\cite{blais2021circuit}.
\section{Results}
\subsection{Purcell Filter Design Considerations}
\begin{figure*}[!htb]
\begin{center}
\includegraphics[width=0.85\twocolfig]{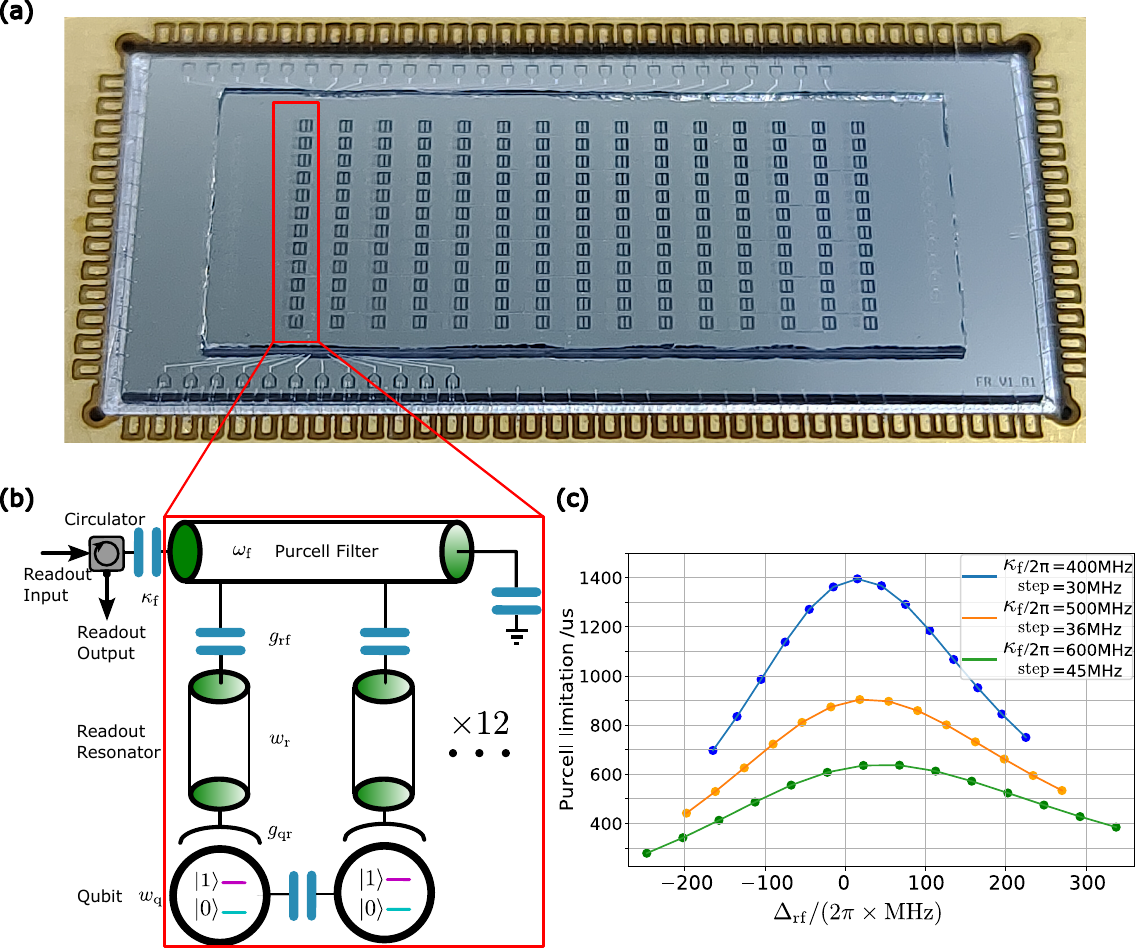}
\end{center}
\caption{
\textbf{Overview of the large-scale flip-chip under our investigation.} 
(a) Photograph of the flip-chip wire-bonded to a PCB. 
This chip features 12 resonators connected to a single shared Purcell filter, with a total of 15 filters linked reflectively to the readout lines. 
The red box (one column) encompasses a single common Purcell filter and 12 resonator-qubit units. 
Within each column, all 12 resonators possess identical etched length, as illustrated in Fig.~\ref{fig:schematic_simulation}(a).
(b) Circuit schematic of the common Purcell filter. 
The qubit frequency $\omega_{\rm q}$ is set to about $2\pi\times 5\GHz$ and the filter frequency is $\omega_{\rm f}/2\pi= 6.7\GHz$. 
The resonator-filter coupling, $g_{\rm rf}$, is chosen to ensure that the resonator damping rate, $\kappa_{\rm r}$, remains constant at $2\pi\times 2\MHz$. 
To optimize the signal-noise ratio of the readout signal, we ensure that $\kappa_{\rm r}$ is approximately double the cross-Kerr nonlinearity, $2\chi$. 
In turn, this ratio is governed by the qubit-resonator coupling $g_{\rm qr}$.
And $\kappa_{\rm f}$ represents the bandwidth of filters.
(c) Purcell limitations for various configurations.
The step indicates the frequency step of resonators. 
The x-axis displays the detuning between the resonator and filter, given by $\Delta_{\rm rf} = \omega_{\rm r}-\omega_{\rm f}$. 
The Purcell limitation is higher for filters with narrower bandwidths and resonators with smaller frequency steps.
}
\label{fig:problem_setup}
\end{figure*}
The flip-chip under investigation is illustrated in Fig.~\ref{fig:problem_setup}(a). 
The chip is fabricated using a sapphire substrate, on which an aluminum film is deposited and subsequently etched to form a specific circuit pattern. 
For details on the fabrication procedure, see Appendix C.
In this chip, 12 quarter-wavelength resonators on the top chip are coupled to a single common Purcell filter~\cite{jeffrey2014fast} situated on the bottom chip.
And a total of 15 filters with specific resonator designs are reflectively connected to the readout lines, as illustrated in Fig.~\ref{fig:problem_setup}(b).
To accommodate an growing number of qubits and resonators with limited readout channels, it is essential to maximize the number of resonators coupled to a single common filter.
In the calculations presented in Appendix A, we demonstrate that the filter bandwidth $\kappa_{\rm f}$ should be smaller to achieve a higher Purcell limitation of the qubit lifetime.
Thus, to maintain a long qubit lifetime in a large-scale qubit chip with filters, the frequency step of the resonators coupled to a shared common Purcell filter should be minimized.
However, reducing the frequency step of resonators with relatively large frequency uncertainty might result in frequency crowding issues.
Therefore, it is crucial to address the frequency uncertainty caused by the flip bonding process.

In the chip we examined, the resonator frequency step is approximately 30\MHz, and the filter bandwidth is set to 600\MHz\ to prevent frequency misalignment between the filter and resonators. 
Under these conditions, the lowest estimated Purcell limitation is around 400\us, as the green curve shown in  Fig.~\ref{fig:problem_setup}, which could be improved in the future by employing a narrower bandwidth filter. 
In the subsequent section, we will present an innovative design to address the frequency uncertainty arising from the flip bonding process.
\subsection{Resonator Design and Simulation}
\begin{figure*}[!htb] 
\begin{center} 
\includegraphics[width=0.9\textwidth]{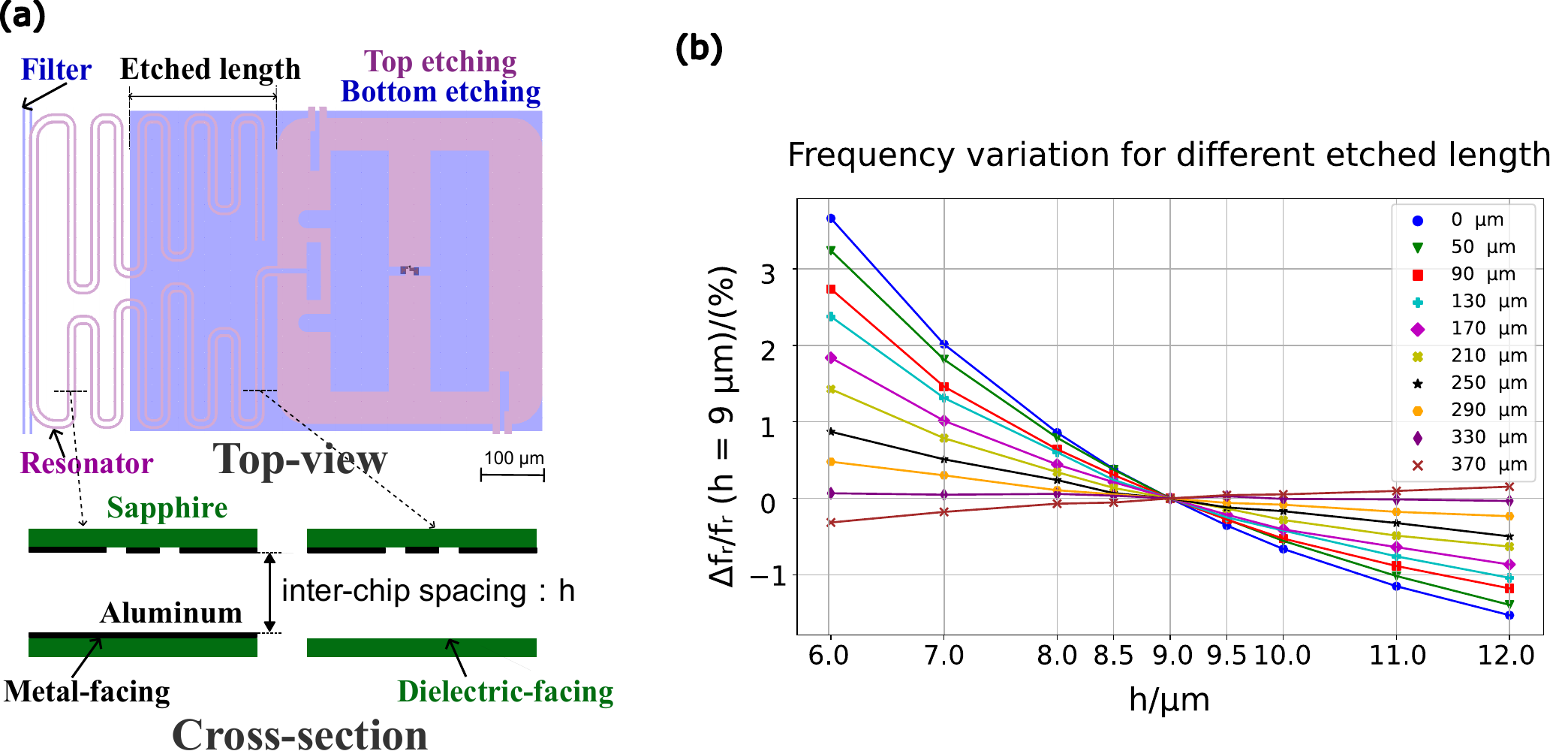} 
\end{center} 
\caption{
\textbf{Schematic of our flip-chip design and the resonator frequency simulation results.}  
(a) Top-view and cross-section of our flip-chip design. 
The chips utilize sapphire as the substrate, which is coated with an aluminum film. 
In the top-view diagram, the blue (pink) area represents the etched portion of the bottom (top) chip, while the aluminum part appears transparent.
In the cross-section diagram, we etch away part of the aluminum film in the region of the bottom chip facing the resonator, exposing the sapphire substrate (dielectric-facing), while the remaining area retains the complete aluminum film (metal-facing). 
The inter-chip spacing between the top and bottom chips is denoted as h.
The region of aluminum film removal is controlled by the ``etched length," as depicted in the top-view diagram.
(b) Dependence of resonator frequencies on inter-chip spacings as determined by electromagnetic simulation.
The x-axis displays the inter-chip spacing h, using 9\um\ as the reference.
The y-axis represents the percentage of frequency change for resonators with varying h, relative to the reference frequencies at h=9\um.
Various curves illustrate different etched lengths.
It is worth noting that for different etched lengths, the reference frequency at h=9\um\ also exhibits minor variations but is designed to remain approximately $6.5\GHz$.
The results show that dielectric-facing and metal-facing have opposing effects on resonator frequency when h deviates from the reference value, and equilibrium is achieved when the etched length is set to 330\um.
} 
\label{fig:schematic_simulation} 
\end{figure*}
The schematic of each individual resonator and its corresponding top and bottom chip is shown in Fig.~\ref{fig:schematic_simulation}. 
The upper part of the Fig.~\ref {fig:schematic_simulation}(a) is a top-view diagram, where the blue (pink) part represents the etched part of bottom (top) chip. 
The resonator and filter are located on the top and bottom chips, respectively, and are coupled to each other.
The lower part of Fig.~\ref{fig:schematic_simulation}(a) shows a cross-section diagram. 
In the unoptimized metal-facing flip-chip design, the resonator frequency is highly sensitive to variations in the inter-chip spacing h. 
In contrast to the metal-facing design, we carried out additional etching of the aluminum film facing the resonator on the bottom chip with a specific length. 
This approach results in the resonator frequency displaying opposite responses to changes in inter-chip spacings between metal-facing and dielectric-facing configurations~\cite{Li2023-ez}.
By partially removing the metal layer facing the resonator, we can effectively mitigate the impact of chip spacing variations on frequency, leading to a  frequency robust design.
Leveraging this mechanism, we designed multiple sets of resonators with differing metal etching lengths. 
Within each set, resonators with the same etching length are designed to couple to a common filter, with their frequencies evenly distributed within the working bandwidth of the filter.
For additional theoretical analysis and calculations, please refer to Appendix F.

Fig.~\ref{fig:schematic_simulation}(b) presents the electromagnetic simulation results of the resonator design.
We set the reference distance for the inter-chip spacing h as 9\um\ and display the percentage change in frequency relative to the reference frequency at h=9\um.
We compare the frequency response of different etching lengths to the h variation. 
As shown in the figure, when the design is not optimized, a 0.8\um\ fluctuation around 9\um\ in chip spacing results in a 70\MHz\ frequency fluctuation of the resonator. 
Considering that the resonator frequency step is only 30\MHz\ and the working bandwidth of our filter is designed to be several hundreds of MHz, such fluctuations can cause significant crosstalk between different resonators. 
As anticipated, partially removing the metal layer opposite the resonator, especially when the etching length reaches an optimal value, leads to a substantial reduction in the variations in the resonator frequency caused by fluctuations in chip spacings. 
Our flip-chip characterization further verifies this result.
\subsection{Chip Spacing Characterization}
\begin{figure*}[!htb] 
\begin{center} 
\includegraphics[width=0.8\textwidth]{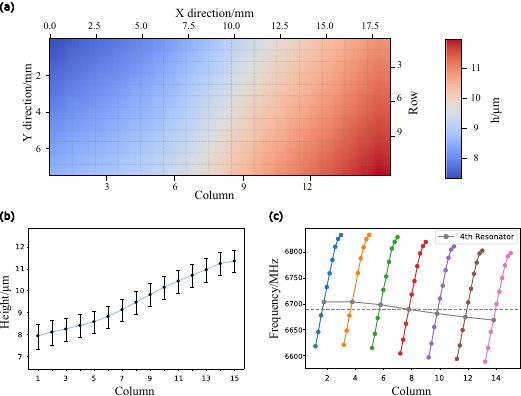} 
\end{center} 
\caption{
\textbf{Measured chip spacings and resonator frequency distribution in the full metal-facing design.}   
(a) Profilometer diagram of chip spacing distribution.
In the X direction, the entire chip is divided into 15 columns, while in the Y direction, it is divided into 12 rows. 
Each grid-enclosed sub-area contains the footprint of a single resonator.
In the X direction, eight groups of resonators with varying etched lengths are arranged in odd-numbered columns, while seven groups of resonators with zero etched length are positioned in even-numbered columns for comparison. 
In the Y direction, the first ten resonators of each group are designed with the same frequency step and will be utilized to evaluate the performance of the resonator design.
(b) Inter-chip spacing distribution for each column measured in room temperature. 
Each data point represents the mean value of the spacing for different resonators within each column, with the bars indicating the range of spacing distribution, which is similar across various columns.
(c) The measured resonator frequency distribution for even-numbered columns.
Since all seven columns share the same design, their frequencies are used to characterize inter-chip spacing fluctuations.
Ideally, all these lines should exhibit uniform linearity.
The gray solid line provides a horizontal comparison of the frequencies for the fourth resonators in each even-numbered column and should maintain a horizontal orientation.
As observed in the figure, the frequency distribution is significantly affected by the tilted spacing of the flip-chip.
} 
\label{fig:Height data} 
\end{figure*}
To assess the spacing between the top and bottom chips, we utilized a stylus profilometer (Dektak XT).
This device directly measures the distance from the top surface of the upper chip to the top surface of the lower chip. 
By subtracting the thickness of the top chip, as determined by the profilometer data, we were able to calculate the spacing between the two chips. 
The top chip was made from double-polished sapphire, and its thickness, which was found to be uniformly distributed at 430$\pm$1\um across the entire chip, was measured using a Scanning Electron Microscope (SEM). 
For more detailed measurement information, please refer to Appendix E.
For a more intuitive representation of the data, we used the upper surface of the bottom chip as the reference plane and extracted the position range of the resonators, resulting in the data shown in Fig.~\ref{fig:Height data}(a).
The chip has a spacing of 9.6$\pm$2.2\um\ progressively increases from the top left to the bottom right and the tilt angle between the top and bottom chips is estimated to be around 219\urad.

The distribution of inter-chip spacings for each column of resonators with different etched lengths is illustrated in Fig.~\ref{fig:Height data}(b). 
To ensure similar variations in spacings for each column, we deliberately controlled the fabrication process. 
Furthermore, we randomized the order of columns with different etched lengths, ensuring that the measured chip frequency distribution is directly related to the etched length and not the specific inter-chip spacing.

Fig.~\ref{fig:Height data}(c) presents the frequency distribution of the seven even-numbered columns, which have identical designs and include resonators with zero etched length. 
These columns are used to characterize the inter-chip spacing fluctuations throughout the entire chip. 
The same color line represents each column, and the intended frequency difference between adjacent resonators is 30\MHz. 
In an ideal scenario, these lines would be linear; however, in practice, a noticeable curvature is observed. 
Each row of resonators is designed to have the same frequency, and ideally, the lines within the same row should be horizontal and straight. 
Nevertheless, a certain degree of inclination is present in reality. 
The curved lines depicted in the figure signify that the resonant frequency substantially deviates from the design value due to the inter-chip spacing tilt.
\subsection{Resonator Optimization Results}
\begin{figure*}[!htb] 
\begin{center} 
\includegraphics[width=0.8\textwidth]{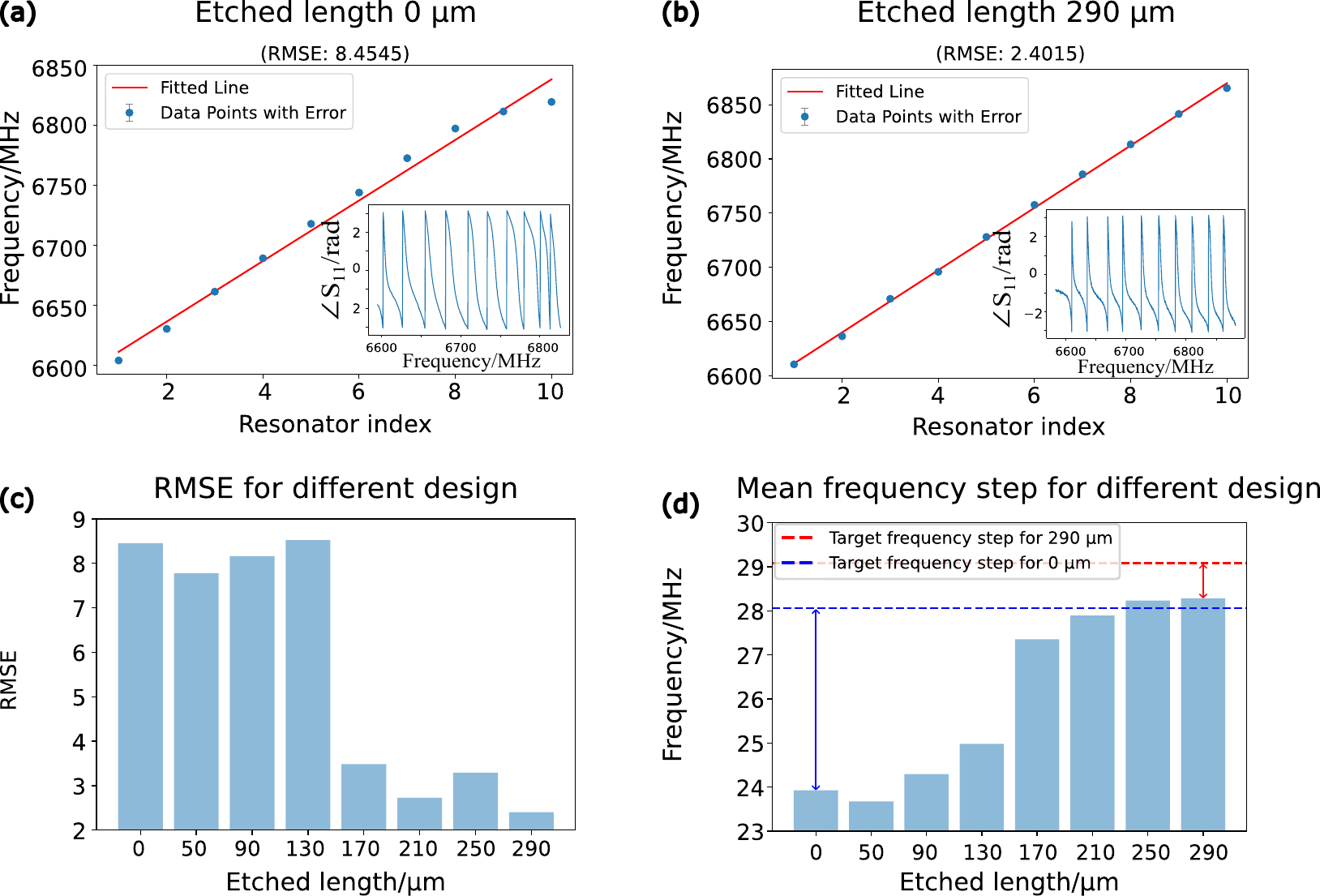} 
\end{center} 
\caption{
\textbf{Comparison of Resonator Frequency Measurements Across Different Etching Lengths.}  
(a, b) The measured resonator frequencies (blue points) and their corresponding linear fittings (red lines) are shown for resonators with etched lengths of 0\um\ and 290\um, respectively. 
The error bars are relatively small and not clearly visible. 
The subplots illustrate the phase response of the reflection coefficient S$_{11}$, which is used to determine the resonator frequencies. RMSE, or Root Mean Square Error, represents the square root of the average squared discrepancies between the predicted and actual values. 
It will be used to assess the linearity of the frequency distribution of the resonator frequencies with various etched lengths. 
(c) RMSE of the measured resonator frequencies for all resonators with differing etched lengths. 
Notably, the design incorporating a 290\um\ etched length demonstrates that the fluctuation of resonator frequencies is small when considering variations in inter-chip spacings. 
(d) Comparison of the average resonator frequency steps, as measured by the slopes of the linear fittings. 
The complete experimental data of the resonator frequency can be found in Appendix D. 
The blue and red dashed lines represent the target resonator frequency steps for etched lengths of 0\um\ and 290\um, respectively, as determined from electromagnetic simulations with an inter-chip spacing of 9\um. 
It is evident that the off-target error of the frequency steps for the 290\um\ design is substantially smaller compared to the unoptimized 0\um\ design.
} 
\label{fig:experiment data} 
\end{figure*}
Ideally, as each column is designed to maintain a consistent frequency step for the first ten resonators, their frequencies should display a uniformly linear distribution.
However, due to varying inter-chip spacings among different resonators, this linear characteristic is distorted. 
Enhanced linearity implies maximizing the utilization of the filter bandwidth, prompting us to optimize the resonator etch length to minimize the frequency congestion. 
We measure the reflected signal from the resonator using a vector network analyzer (VNA) and obtain the resonator frequency through fitting, as described in Ref.~\cite{mcrae2020materials}.
All the measured data points in Fig.~\ref{fig:experiment data}(a,b) are the average of 30 measurements with $-140\dBm$ input power at the chip. 
The VNA bandwidth is set to be 5\kHz. 
To assess the performance of our design optimization, we present the linear fittings (obtained using the least squares method) based on the experimentally measured resonator frequency data, as shown in Fig.~\ref{fig:experiment data}(a,b). 
We then compare the linearity of the frequency distribution for each column using the root mean square error (RMSE), defined as RMSE $= \sqrt{\frac{1}{n} \sum_{i=1}^n (y_i - \hat{y}_i)^2}$, where $(y_i - \hat{y}_i)^2$ represents the squared difference between the observed value $y_i$ and the predicted value $\hat{y}_i$ for each data point i, and n is the total number of data points.

We employ RMSE as a measure of the linearity of the resonator frequency distribution, with smaller values being more favorable. 
A comparison of the RMSE for various etching proportions can be found in Fig.~\ref{fig:experiment data}(c). 
Notably, the linearity of the optimized partial dielectric-facing design has improved by a factor of more than 3.5 compared to the metal-facing design. 
The theoretically optimal etched length of 330\um\ is not presented due to fabrication damage on the filter. 
Furthermore, we exclude the last two resonators in each column from our analysis, as they are designed with different shapes.

As shown in Fig.~\ref{fig:experiment data}(d), we utilize the slope of the linear fitting to represent the mean frequency step. 
The dashed lines represent the simulated design target frequency steps. 
We find that, when the etched length is 290\um, the off-target error of the measured frequency step relative to the design value has been optimized to less than 1/5 of the original metal-facing design. 
Our experimental results confirm that the partial dielectric-facing design maintains precise resonator frequency control despite the chip spacing fluctuations depicted in Fig.~\ref{fig:Height data}(a).
\section{Discussion}
We have shown that the new resonator design can improve the accuracy of the resonator frequency target in the flip-chip architecture. 
However, we also discovered that achieving optimal results requires etching a large area of the bottom chip metal. 
This etched area accommodates a mode with a specific frequency and loss. 
In the optimized design, the etched length is long, the additional mode frequency is low, and the etched area is close to the filter, which is capacitively coupled to the $50 \Omega$ environment. 
It is crucial to investigate whether this additional mode will cause significant leakage of the qubit.

In the designed resonator with an etched length of 330\um, the simulated additional mode frequency is $\omega_{\rm etch}/2\pi = 14.8\GHz$. 
Based on the $50 \Omega$ impedance boundary condition of the filter, the coupling quality factor $Q_{\rm c}$ of this mode to the environment is approximately 14000. 
The coupling strength between the qubit and the additional mode, $g_{\rm q-etch}/2\pi$, is estimated to be 96\MHz\ using the EPR method~\cite{minev2021energy}. 
With these parameters and considering a 5\GHz\ qubit, the Purcell limit through this additional mode is estimated to be approximately 14.5\ms, which can be safely ignored. 
In our experiments, we have also tested 18 qubits with lifetimes ranging from 17.2\us\ to 106.7\us\ and an average of 52.6\us. 
These results are consistent with the current manufacturing standards for aluminum film flip-chip. 
Therefore, the new resonator design does not compromise the qubit coherence.
\section{Conclusion}
In conclusion, we have presented an innovative resonator design within the flip-chip architecture that maintains frequency stability despite inter-chip spacing fluctuations.
This development signifies a considerable advancement towards improving the performance of multi-qubit structures. 
The same principle can be extended to other quantum components based on the coplanar-waveguide (CPW) resonator structure within the flip-chip architecture, such as the Purcell filter and the qubit coupling bus~\cite{majer2007coupling}. 
Additionally, it can be applied beyond the field of superconducting qubits, for instance, in potential applications within the quantum dot system featuring the flip-chip architecture~\cite{holman20213d,corrigan2023longitudinal}.

\section{Appendix}
\subsection{Appendix A: Purcell Limitation}
Here, we present comprehensive calculations of the Purcell limitation as a complement to the theoretical study in reference~\cite{sete2015quantum}. 
We begin with a qubit-resonator combined system that does not include a Purcell filter.
Within the one-excitation subspace of the Hilbert space, the Hamiltonian of this combined system can be represented as: 
\begin{equation}
    H_{\rm qr}=\begin{pmatrix}
      \omega_{\rm q} & g_{\rm qr} \\
      g_{\rm qr}^{*} & \omega_{\rm r}-i\kappa_{\rm r}/2
    \end{pmatrix},
    \label{eq:H_qr}
\end{equation}
where $\omega_{\rm q}=2\pi\times 5\GHz$ denotes the qubit frequency, $\omega_{\rm r}=2\pi\times 6.7\GHz$ represents the resonator frequency, $g_{\rm qr}$ indicates the coupling strength between the qubit and the resonator, and $\kappa_{\rm r}=2\pi\times 2\MHz$ signifies the dissipation rate to the environment.

To achieve a high signal-to-noise ratio for the readout signal, we set $\kappa_{\rm r}$ equal to twice the cross-Kerr nonlinearity, $2\chi$, where $\chi \simeq -g_{qr}^2 / (\Delta_{\rm qr} + \Delta_{\rm qr}^2 / \eta)$~\cite{blais2021circuit}, $\Delta_{\rm qr}=\omega_{\rm q}-\omega_{\rm r}$ represents the detuning between the qubit and the resonator, and $\eta= -2\pi\times 270\MHz$ denotes the self-Kerr nonlinearity of the qubit. 
Under this constraint, we can easily obtain $g_{\rm qr} = 2\pi\times 111.37\MHz$. 
Once all parameters are determined, we can diagonalize Eq.~\ref{eq:H_qr} and obtain the complex qubit mode eigenfrequency $\widetilde{w}_{\rm q}$. 
The Purcell limitation of the qubit can then be calculated as $T_1 = 1/(-2\rm{Im}(\widetilde{w}_{\rm q})) \simeq 18.78\us$. 
It is evident that, without a Purcell filter, the qubit coherence would be significantly affected by leakage into the environment through the resonator.

In the same way, we can calculate the Purcell limitation after adding the Purcell filter, which is a lossy resonator. 
The Hamiltonian of the qubit-resonator-filter combined system can be represented as:
\begin{equation}
    H_{\rm qrf}=\begin{pmatrix}
      \omega_{\rm q} & g_{\rm qr} &0 \\
      g_{\rm qr}^{*} & \omega_{\rm r} & g_{\rm rf}\\
      0&g_{\rm rf}^{*} & \omega_{\rm f}-i\kappa_{\rm f}/2\\
    \end{pmatrix},
    \label{eq:H_qrf}
\end{equation}
where $\omega_{\rm f} = 2\pi\times 6.7\GHz$ denotes the filter frequency, $\kappa_{\rm f}$ is the coupling strength between the filter and the environment, and $g_{\rm rf}$ indicates the resonator-filter coupling strength. 
In the calculation, we first assume a filter bandwidth $\kappa_{\rm f}$. 
Then, we diagonalize $H_{\rm qrf}$, obtain the complex resonator mode eigenfrequency $\widetilde{w}_{\rm r}$, and find $g_{\rm rf}$ by setting $ -2\rm{Im}(\widetilde{w}_{\rm r})= \kappa_{\rm r}= 2\pi\times 2\MHz$. 
Once $g_{\rm rf}$ is determined and since $g_{\rm qr}$ is already known, we can quickly find $\widetilde{w}_{\rm q}$, and thus the Purcell limit of qubit coherence, given by $T_1 = 1/(-2\rm{Im}(\widetilde{w}_{\rm q}))$.

To determine the optimal parameters for the filter and resonators, we initially set the damping rate of the resonators $\kappa_{\rm r} = 2\pi\times 2\MHz$ and the frequency of the filter $\omega_{\rm f} = 2\pi\times 6.7\GHz$. 
We then explore various filter bandwidths $\kappa_{\rm f}$ to evaluate the Purcell limitations of the qubit mode lifetimes for resonators with different linear frequency differences, as indicated by `step' in Fig.~\ref{fig:problem_setup}(c). 
For each configuration, we consider 14 resonators, and their detuning to the filter $\Delta_{\rm rf} = \omega_{\rm r}-\omega_{\rm f}$ is displayed as the x-axis of Fig.~\ref{fig:problem_setup}(c).
As evident in Fig.~\ref{fig:problem_setup}(c), the Purcell limitation is higher for filters with narrower bandwidths and resonators with smaller frequency steps.
\subsection{Appendix B: Experimental Setup}
\begin{figure*}[!htb] 
\begin{center} 
\includegraphics[width=0.5\textwidth]{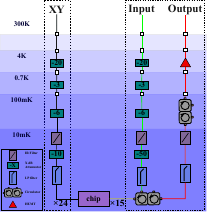} 
\end{center} 
\caption{
\textbf{Full wiring diagram for the experimental setup of the tested chip.}  
The experimental schematic consists of three unique signal channel types: 24 black XY channels, 12 green input, and 12 red output readout channels. 
The XY and readout signals interface with room temperature electronics through an array of components such as filters, attenuators, and circulators. 
On the room temperature side of the output line, a 40\dB\ microwave amplifier is positioned, which is not depicted in this illustration.
} 
\label{fig:cryo} 
\end{figure*}
We mount the 180-resonator flip-chip at the base plate of a dilution refrigerator and connect it to the control electronics setup located at room temperature, as shown in Fig.~\ref{fig:cryo}. 
Lines for XY control (black), readout input (green), and readout output (red) are configured with the indicated microwave components for signal conditioning. 
The readout input/output lines are connected to a VNA to determine the resonator frequencies. 
\subsection{Appendix C: Fabrication}
The flip-chip is divided into two parts: the top chip (25 mm by 10 mm) and the bottom chip (30 mm by 13 mm). 
The readout resonator is located on the top chip, while the Purcell filter and readout transmission line are situated on the bottom chip. 
The top and bottom chips are connected by indium bumps, providing both mechanical support and electrical connection. 
To prevent the indium bumps from interdiffusing with the underlying aluminum metal and forming brittle intermetallic compounds, a layer of niobium serves as an under bump metal (UBM) layer between the indium bumps and aluminum. 
To facilitate the measurement of the spacing between the top and bottom chips, the top chip employs double-polished sapphire, and the bottom chip uses single-polished sapphire.
First, we employ the e-beam method to deposit aluminum with a background vacuum of E-9 torr. 
Second, we utilize ultraviolet exposure to define the resonators on the top chip and the transmission lines on the bottom chip. 
Then, we apply wet etching type A (TRANSENE COMPANY, INC.) to transfer the pattern to the underlying metal. 
Third, we expose the UBM layer pattern on the double-layer resist of LOR (MicroChem)and S1805 (DOW), deposit niobium by magnetron sputtering, and obtain the required UBM structure using a stripping process.
Fourth, we fabricate full-coverage and discrete air bridges on the bottom chip. 
For full-coverage air bridges, we use periodic opening windows to achieve stable support and effective degumming. 
Fifth, we expose the negative resist on the top and bottom chips, deposit 9\um\ of indium by thermal evaporation, and obtain indium solder bumps using a stripping method. 
Sixth, we employ a dicing machine to cut the crystal grains.
Finally, we flip the top chip, use an inverted soldering machine (SET, FC150) to adjust the parallelism and planar alignment of the top and bottom chips, and press the top and bottom chips together at room temperature with a pressure of 2\kg. 
After press-welding, the chip is wire-bonded onto a PCB board and enclosed in an aluminum sample box.
\subsection{Appendix D: More Experiment Data}
\begin{figure*}[!htb] 
\begin{center} 
\includegraphics[width=1\textwidth]{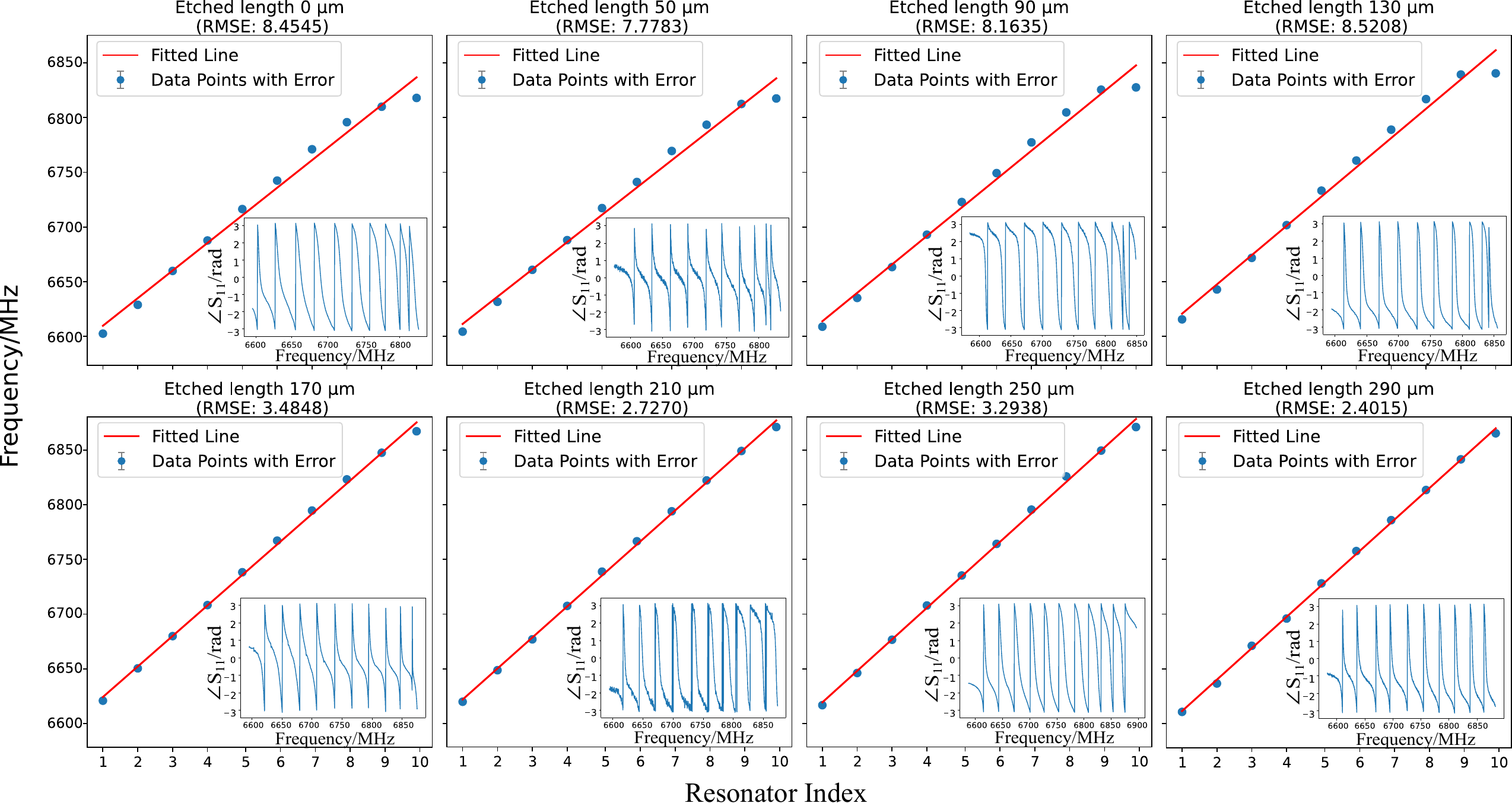} 
\end{center} 
\caption{
\textbf{All resonator frequency measured from the experiment.}  
The subplots display the full set of experimentally measured resonator frequencies for different etched lengths, ranging from 0\um\ to 290\um. The red lines represent the corresponding fitted lines, and the inset shows the corresponding S$_{11}$ plot.
} 
\label{fig:appendix_D} 
\end{figure*}

\subsection{Appendix E: SEM Measurement}
\begin{figure*}[!htb] 
\begin{center} 
\includegraphics[width=0.8\textwidth]{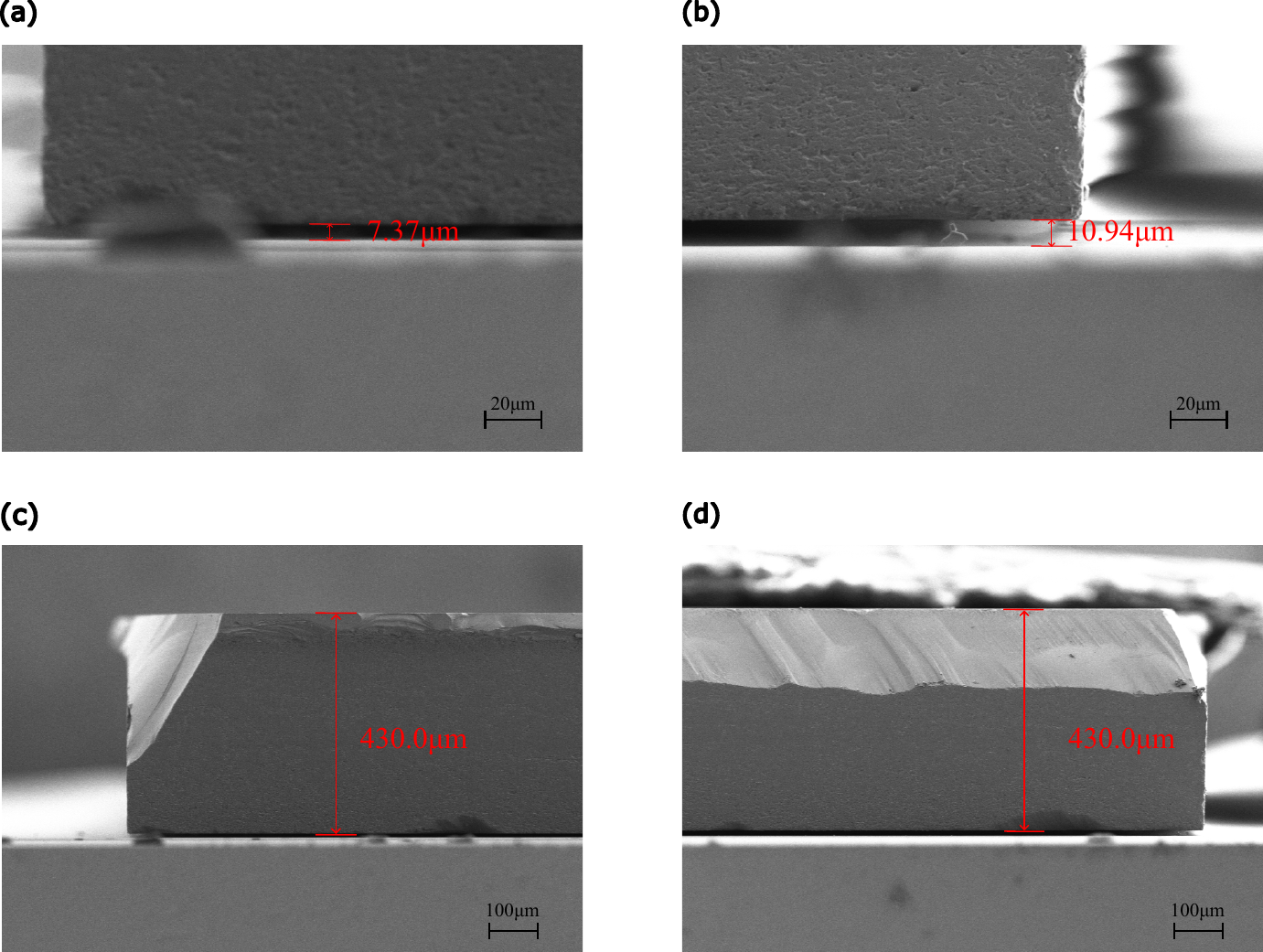} 
\end{center} 
\caption{
\textbf{SEM micraograph of the flip-bonding device.}  
Thickness measurements of the top chip at the upper-left edge (a) and upper-right edge (b). 
Inter-chip spacings measurements at the upper-left edge (c) and upper-right edge (d).}
\label{fig:SEM micraograph of the flip-bonding device.} 
\end{figure*}
The height data acquired by the profilometer represents the sum of the top chip's thickness and the inter-chip spacing. 
Hence, precise measurement of the top chip's thickness is crucial for determining the inter-chip spacing. 
After completing the low-temperature measurements, we  employed SEM (ZEISS, Gemin) to assess the thickness at various positions of the top chip. 
The results revealed a highly uniform thickness of 430\um, as depicted in Fig.~\ref{fig:SEM micraograph of the flip-bonding device.}(c,d).

Moreover, we directly inspected the inter-chip spacing between the top and bottom chips using SEM. 
Due to the disparity in height between the edges of the top and bottom chips, simultaneous clear capture by SEM was challenging. 
To prevent the bottom chip's circuit structures from obscuring the top chip, we deliberately tilted the sample by 1 degree, resulting in a measurement error of approximately a hundred nanometers. 
The inter-chip spacing at various positions is depicted in Fig.~\ref{fig:SEM micraograph of the flip-bonding device.}(a,b). 
SEM results showed a spacing difference of 3.57\um\ between the top left and top right corners of the chip, aligning with the height difference of 3.26\um\ measured by the profilometer. 
This cross-validation suggests that the method of using a profilometer to measure the inter-chip spacing and SEM to measure the top wafer's thickness, thereby determining the spacing distribution, is reliable.

\subsection{Appendix F: Theoretical Analysis of Resonator Frequency}
\begin{figure*}[!htb] 
\begin{center} 
\includegraphics[width=1\textwidth]{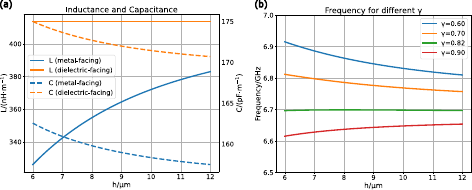} 
\end{center} 
\caption{
\textbf{Analysis of the impact of dielectric-facing and metal-facing on the resonator frequency.}
(a) Capacitance and inductance dependence on inter-chip spacings for metal- and dielectric-facing configurations.
(b) Theoretical predictions of resonator frequency for different etching ratios.
Each curve shares the same effective length of 4.5 \mm\ for CPW and varies only in terms of the overall etching ratio $\gamma$.
For the optimized $\gamma=0.82$, the resonator frequency will be insensitive to a large inter-chip spacing fluctuation.
} 
\label{fig:theory} 
\end{figure*}

In this section, we offer a comprehensive explanation of the theoretical basis for attaining resonator frequency insensitivity to inter-chip spacing variations using the partial etching method implemented in our study.

The resonator frequency is inversely proportional to the square root of the product of capacitance and inductance. 
Inductance can be categorized into two parts: geometric and kinetic. 
Since kinetic inductance is significantly smaller than geometric inductance, and there is a notable difference between the geometric inductance in dielectric-facing and metal-facing configurations, we exclude kinetic inductance from our discussion on the optimal etching ratio and the impact of inter-chip spacing variations on resonator frequency.
To quantify the effect of chip spacings on the resonator frequency, we utilized the capacitance and inductance formulas for CPW in the flip-chip~\cite{simons2004coplanar} and plotted the variations of these parameters with inter-chip spacings in Fig.~\ref{fig:theory}(a). 
The curves show that the resonator capacitance decreases in both configurations as the chip spacing increases. 
However, in terms of inductance, we observe that it increases with enlarging chip spacings in the metal-facing configuration, while remaining unaffected by chip spacings in the dielectric-facing scenario.
This phenomenon can be attributed to the fact that in the dielectric-facing scenario, the current flows exclusively on the top chip, making the inter-chip spacing irrelevant to the current distribution. Conversely, in the metal-facing configuration where both sides are metallic, the current flow distribution is influenced by the metal presence on both sides.

Upon further analysis, we find that the metal-facing configuration exhibits a higher rate of change in inductance with inter-chip spacings compared to capacitance. 
As a result, frequency variations in the metal-facing configuration are primarily influenced by inductance, while those in the dielectric-facing configuration are primarily influenced by capacitance. 
We define the overall etching ratio $\gamma$ as the length ratio of the dielectric-facing part to the entire length of the CPW. 
Through analytical calculations, we determine that the optimized value of $\gamma$ in our design is approximately 0.82, as depicted in Fig.~\ref{fig:theory}(b). 
This finding aligns with both our electromagnetic simulations and experimentally measured optimal etching ratios.

\ \\
\textbf{Data availability}
The source data for this paper can be requested from the corresponding author with a reasonable justification.\\
\textbf{Code availability}
Codes used in the theoretical calculation are available from the corresponding author on reasonable request.\\
\bibliography{MRF}
\newpage
\section{Acknowledgments}
We express our appreciation to T.Q. Cai for providing the sample box, and to M.C. Dai for offering valuable insights regarding room temperature chip measurements. 
Furthermore, we acknowledge the dedicated efforts of the electronics team at Tencent Quantum Lab in preparing the room-temperature electronics.
\section{Author contributions}
S.M.A. conceptualized the experiment. 
Chip design and simulation were carried out by S.M.A., T.H.W., J.J.H., and D.F.L.
Y.L. was responsible for device fabrication and conducted room temperature measurements. 
Low-temperature microwave measurements were performed by S.M.A. and D.F.L.
Data analysis and manuscript preparation were completed by T.H.W., S.M.A., and Y.L., with contributions from all authors. 
\section{Competing Interests Statement}
The authors declare no competing interests.
\newpage
\end{document}